\documentclass[jounral]{IEEEtran}

\ifCLASSINFOpdf
\else
   \usepackage[dvips]{graphicx}
\fi
\usepackage{url}

\usepackage{cite}
\usepackage{graphicx,amsmath,amssymb,tikz,psfrag}
\usepackage{color}
\usepackage{fancyhdr} 
\usepackage{tikz}
\usetikzlibrary{shapes,arrows,chains}
\usepackage{xcolor}
\usepackage{algorithm}
\usepackage{algpseudocode}
\usepackage{enumerate}

\begin{document}

\title{Secure Degrees-of-Freedom of the MIMO X Channel with Delayed CSIT}

\author{Tong~Zhang, and Rui~Wang
	\thanks{		
	The authors are with the University Key Laboratory of Advanced Wireless Communications of Guangdong Province, Department of
	EEE, Southern University of Science and
	Technology, Shenzhen 518055, China (email: bennyzhangtong@yahoo.com,
	and wang.r@sustech.edu.cn).
}}
 
\maketitle

\begin{abstract}
In this paper, we study the secure degrees-of-freedom (SDoF) characterization for the multiple-input multiple-output (MIMO) X channel with confidential messages and delayed channel state information at the transmitter (CSIT).  In particular, we propose a transmission scheme, which can be regarded as a generalization of the state-of-the-art scheme without security and with delayed CSIT. The key of this generalization is performing the security analysis, by which we derive the optimal duration of the artificial noise transmission phase. As a result, we derive the sum-SDoF lower bound. Furthermore, we reveal that if the number of receive antennas, denoted by $N$, is fixed, the minimum number of transmit antennas achieving the maximum of the lower bound is $\frac{7+\sqrt{33}}{8}N$.
\end{abstract}

\begin{IEEEkeywords}
Delayed CSIT,  information-theoretic security, lower bound, MIMO X channel, secure degrees-of-freedom. 
\end{IEEEkeywords}

\IEEEpeerreviewmaketitle

\section{Introduction}

\IEEEPARstart{T}{he} degrees-of-freedom (DoF) characterization for the  multiple-input multiple-output (MIMO) X channel with delayed channel state information at the transmitter (CSIT) has attracted lots of interests \cite{71,72,107,64}.  In \cite{71}, a non-trivial sum-DoF lower bound was achieved by a novel transmission scheme for the single-input single-output (SISO) X channel with delayed CSIT. Since each transmitter has messages for all receivers, the scheme in \cite{71} for X channel is different from the schemes for broadcast channel \cite{00}  and interference channel \cite{01}. This transmission scheme was shown to be linear sum-DoF optimal in \cite{72}. Thereafter, in \cite{107}, the transmission scheme in \cite{71} was generalized to the MIMO X channel with delayed CSIT. However, the study of \cite{64} showed that the general transmission scheme in \cite{107} was linear sum-DoF optimal, except one antenna configuration case. For this case, a linear sum-DoF optimal transmission scheme was proposed in \cite{64} to fill the gap. Unlike the no delayed CSIT utilization for data transmission phase in \cite{107}, for this antenna configuration case, the scheme in \cite{64} exploits the delayed CSIT in one transmitter when the other transmitter is sending data symbols. 

The secure degrees-of-freedom (SDoF) region of MIMO interference channel  with confidential messages (ICCM) was characterized in \cite{4}. The sum-SDoF of SISO X channel with confidential messages (XCCM) was studied in \cite{301,302}. 
The research of SDoF with delayed CSIT was stemmed from \cite{31}, where the SDoF region for two-user MIMO broadcast channel with confidential messages (BCCM) was characterized. In \cite{31}, the key idea of the transmission scheme is to add an artificial noise (AN) transmission phase before the data transmission phase. For MIMO ICCM with delayed CSIT, a sum-SDoF lower bound  was proposed in \cite{35}. For MIMO XCCM with delayed CSIT and output feedback, the SDoF region was derived in \cite{32}. With alternating no, delayed, and current CSIT, the SDoF region of two-user multiple-input multiple-output (MISO) BCCM was characterized in \cite{111}. Under no eavesdropper's CSIT, the sum-SDoF of one-hop wireless networks were obtained in \cite{300}.
However, there is no research explore the SDoF of the MIMO XCCM with delayed CSIT, which is the focus of this paper. 

The main contribution of this paper is that we obtain a non-trivial sum-SDoF lower bound by designing a  transmission scheme. Our transmission scheme cannot be covered by those schemes in \cite{31,35,32} and their extensions.  
Instead, the proposed transmission scheme can be regraded as a generalization of the scheme in \cite{64} for symmetric antenna configurations, since the security issue is considered by us. To generalize the scheme in \cite{64}, we first add an AN transmission phase before data transmission phase.  Next, the transmitted data symbols are masked with feedback received AN signals, where the arrangement of data transmission mimicks that in \cite{64}. However, this raises a problem: What's the optimal duration of the AN transmission phase? We answer this question by performing the security analysis. Similar to the security analysis in \cite{31,35,32}, we apply data processing inequality and Lemma 2 in \cite{31} to transform the mutual information expression for information leakage into matrix rank expressions. Whereas, since the delayed CSIT setting for XCCM is not considered in \cite{31,35,32}, the deduced matrix expressions and their rank analysis are different from that in \cite{31,35,32}. Thus, the derived optimal duration of AN transmission phase is new. Interestingly, our lower bound indicates that if the number of receive antennas is fixed, there is a minimum number of transmit antennas achieving the maximum of the lower bound.

 \textit{Notations}:  The identity matrix of dimensions $m$ is denoted by $\textbf{I}_m$.  The determinant of matrix $\textbf{A}$ is denoted by $\det(\textbf{A})$. The block-diagonal matrix with blocks $\textbf{\textsc{P}}$ and $\textbf{Q}$ is denoted by $
\text{bd}\{\textbf{P}, \textbf{Q}\} = 
[\textbf{P}, \textbf{0};
\textbf{0},\textbf{Q}]
$. The $\log$ function is referred to $\log_2$.

\section{System Model and Main Results}

\subsection{$(M,M,N,N)$ MIMO XCCM with Delayed CSIT}

We consider a $(M, M, N, N)$ MIMO XCCM has two transmitters with $M$ antennas and two receivers  with $N$ antennas, i.e.,  transmitters 1, 2, and receivers 1, 2. The transmitter $i=1,2$ has a confidential message $W_{i,j}$ for receiver $j=1,2$. The complex input signal at transmitter
$i = 1, 2$ and time slot (TS) $t$ is denoted by $\textbf{x}_i[t] \in \mathbb{C}^M$. The complex
received signal at receiver $j = 1, 2$ and TS $t$ is denoted by
$\textbf{y}_j[t] \in \mathbb{C}^N$. Mathematically, the input-output relationship is written as
\begin{equation}
 \textbf{y}_j[t] = \textbf{H}_{1,j}[t]\textbf{x}_1[t] + \textbf{H}_{2,j}[t]\textbf{x}_2[t] + \textbf{z}_j[t], \quad j=1,2,
\end{equation}
where the CSI matrix from the transmitter $i=1,2$ to the receiver $j=1,2$ at TS $t$ is denoted by $\textbf{H}_{i,j}[t] \in \mathbb{C}^{N \times M}$, and the additive white Gaussian noise (AWGN) vector at the receiver $j$ and TS $t$ is denoted by $\textbf{z}_j[t]$.  We assume that  $\textbf{H}_{i,j}[t],\forall t$ is non-static (time-varying) and linearly independent. We denote the collection of CSI matrices for TS $1$ to TS $t-1$ as $\textbf{H}^{t-1} = [\textbf{H}_{i,j}[1],\cdots,\textbf{H}_{i,j}[t-1]],i,j=1,2$. At TS  $t$, due to feedback delay, $\textbf{H}^{t-1}$ available at  two transmitters.  

\subsection{Sum-SDoF}
A $(2^{nR_{1,1}(\text{SNR})},2^{nR_{1,2}(\text{SNR})}$,$2^{nR_{2,1}(\text{SNR})}$,$2^{nR_{2,2}(\text{SNR})}$,$n)$ code with secure  achievable rates $R_{i,j}(\text{SNR})$, $i,j=1,2$ is defined as follows: The communication process takes $n$ channel uses with confidential messages $W_{i,j}=[1,\cdots,2^{nR_{i,j}(\text{SNR})}],i,j=1,2$.  A stochastic encoder $f_i(\cdot)$ at the transmitter $i=1,2$, encodes confidential message $W_{i,1}$, $W_{i,2}$, and  $\textbf{H}^{t-1},$ to a codeword $\textbf{x}_i^n = [\textbf{x}_i[1],\cdots,\textbf{x}_i[n]]$.  At the TS $t$, the input signal is encoded by $
\textbf{x}_i[t] = f_i(W_{i,1}, W_{i,2}, \textbf{H}^{t-1}), i = 1,2.
$
A decoder $g_{i,j}(\cdot)$ at the receiver $j=1,2$ decodes the output signal $\textbf{y}_j^n \triangleq \{\textbf{y}_j[1],\cdots,\textbf{y}_j[n]\}$ to an estimated message $\widehat{W}_{i,j}$, which is given by
$
\widehat{W}_{i,j} = g_{i,j}(\textbf{H}^n,\textbf{y}_j^n), j =1,2,
$
where two receivers are assumed to have perfect CSI. In addition, the secure code  should satisfy the reliability criterion, i.e.,
$\label{R1}
\Pr[W_{i,j} \ne \widehat{W}_{i,j}] \le \epsilon_n, i,j = 1,2,
$ and the secrecy criterion, 
\begin{subequations}
	\begin{eqnarray}
&&   \frac{1}{n}I(W_{1,1},W_{2,1};\textbf{y}_2^n)  \le \epsilon_n, \label{S1} \\
&&   \frac{1}{n}I(W_{1,2},W_{2,2};\textbf{y}_1^n)  \le \epsilon_n, \label{S2}
	\end{eqnarray}
\end{subequations}
where $\epsilon_n \rightarrow 0$ as $n \rightarrow {\cal{1}}$. 
The secure sum-capacity is defined as the maximal achievable sum-rate, which is written as 
$
C = \max  \,\, \sum_{i=1}^2\sum_{j=1}^2 R_{i,j}(\text{SNR}).
$
The sum-SDoF is a first-order approximation of the secure sum-capacity in the high SNR regime and defined as follows:
\begin{equation}
\sum_{i=1}^2\sum_{j=1}^2 d_{ij} = \lim_{\text{SNR} \rightarrow {\cal{1}}}  \frac{C}{\log \text{SNR}}.
\end{equation}

\subsection{Main Results}
\textbf{Theorem 1}: Consider the $(M,M,N,N)$  MIMO XCCM with delayed CSIT. The sum-SDoF lower bound is given by
\begin{equation} \label{LB}
\sum_{i=1}^2\sum_{j=1}^2 d_{ij}\ge 
\begin{cases}
0, & M \le N, \\
\frac{3N(M-N)}{2M-N},& N < M  \le \frac{7+\sqrt{33}}{8}N, \\
\frac{6MN}{8M-N}, & \frac{7+\sqrt{33}}{8}N < M \le 2N, \\
4N/5, & 2N < M. \\
\end{cases}
\end{equation}
\begin{IEEEproof} 
Please refer to Section-III.
\end{IEEEproof}

\begin{figure}
	\centering
	\includegraphics[width=2.7in]{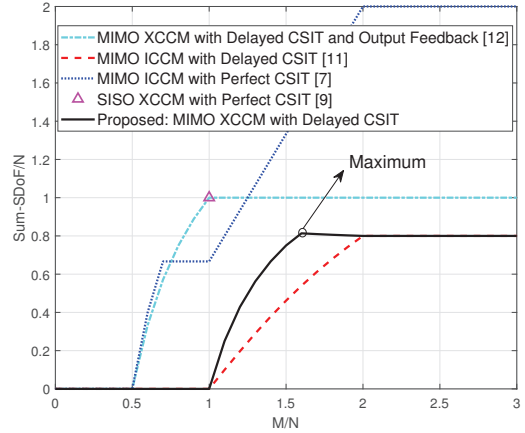}
	\caption{The derived sum-SDoF lower bound is compared with related results.} \label{F1}
\end{figure}

\textit{Remark}:  Fig. \ref{F1}  shows: 1) The derived sum-SDoF lower bound has a gain over the sum-SDoF lower bound of MIMO ICCM with delayed CSI  \cite{35}, where the gain comes from the joint data transmission from two transmitters; 2) The derived sum-SDoF lower bound is less than that of the scenarios with better CSIT conditions, i.e., the sum-SDoF of MIMO ICCM with perfect CSIT \cite{4},  the sum-SDoF of SISO XCCM with perfect CSIT \cite{302}, and the sum-SDoF of MIMO XCCM with delayed CSIT and output feedback \cite{32}; 3) By applying the proposed scheme, the derived lower bound decreases in $(7+\sqrt{33})/8 < M/N \le 2$. This implies that we can switch off $M - (7+\sqrt{33})N/8$ antennas for $(7+\sqrt{33})/8 < M/N$, to make the sum-SDoF lower bound non-decreasing.

\section{Proof of Theorem 1}

\subsection{$M \le N$: Keep Two Transmitters Silent}

Intuitively, the transmitted AN symbols from one transmitter will be immediately decoded by the eavesdropper, which disables the security of data transmission superposed feedback received AN signals. Hence, the sum-SDoF lower bound is $0$.

\subsection{$N < M  \le 2N$: The Proposed Transmission Scheme}

 The following pre-assigned matrices: $\phi[k] \in \mathbb{C}^{M \times \tau_1N}, k=1,\cdots,\tau_2$, $\omega[k] \in \mathbb{C}^{M \times \tau_1N}, k=1,\cdots,\tau_3$, $\gamma[k] \in \mathbb{C}^{M \times \tau_2N}, k=1,\cdots,\tau_3$,
 $\theta[k] \in \mathbb{C}^{N \times \tau_3N}, k=1,\cdots,\tau_4$, are linearly independent and full rank. Holistically, we denote $\Phi = [\phi[1];\cdots;\phi[\tau_2]]$, $\Omega = [\omega[1];\cdots;\omega[\tau_3]]$, $\Gamma=[\gamma[1];\cdots;\gamma[\tau_3]]$, and $\Theta=[\theta[1];\cdots;\theta[\tau_4]]$.

\underline{\textit{Phase-I}} \textit{(AN Symbol Transmission for Receiver 1)}: This phase spans $\tau_1$ TSs. At TS $t=1,\cdots,\tau_1$, $M$ AN symbols  are sent from transmitter $1$, i.e., $\textbf{x}_1^\text{I}[t] = \textbf{u}_1[t]$. Meanwhile, the transmitter $2$ keeps silent.  The holistic transmitted signal for Phase-I is written as
\begin{eqnarray}
&& \textbf{x}_1^\text{I} = \textbf{u}_1.
\end{eqnarray}
The holistic received signals for Phase-I are written as
\begin{eqnarray}
&& \textbf{y}_j^\text{I} = \textbf{H}_{1,j}^\text{I}\textbf{u}_1 + \textbf{z}_j^\text{I},\quad j=1,2,
\end{eqnarray} 
where the AWGN signal at receiver $j$ is denoted by  $\textbf{z}_j^\text{I}$, $\textbf{u}_1 = [\textbf{u}_1[1];\cdots;\textbf{u}_1[\tau_1]]  \in \mathbb{C}^{\tau_1M}$, and  $\textbf{H}_{1,j}^\text{I} = \text{bd}\{\textbf{H}_{1,j}[1],\cdots,\textbf{H}_{1,j}[\tau_1]\} \in \mathbb{C}^{\tau_1N \times \tau_1M}, j=1,2$.

\underline{\textit{Phase-II}} \textit{(AN Symbol Transmission for Receiver 2)}: This phase is same as Phase-I, except the role of the transmitters 1 and 2 is swapped. Hence, this phase spans $\tau_1$ TSs as well. At TS $t=\tau_1+1,\cdots,2\tau_1$, $M$ AN symbols are sent from transmitter $2$, i.e., $\textbf{x}_2^\text{II}[t] = \textbf{u}_2[t-\tau_1]$. Meanwhile, the transmitter $1$ keeps silent. The holistic transmitted signal for Phase-II is written as
\begin{eqnarray}
	&& \textbf{x}_2^\text{II} = \textbf{u}_2.
\end{eqnarray}
The holistic received signals for Phase-II are written as  
\begin{eqnarray}
	&& \textbf{y}_j^\text{II} = \textbf{H}_{2,j}^\text{II}\textbf{u}_2 + \textbf{z}_j^\text{II}, \quad j=1,2,
\end{eqnarray}
where  the AWGN signal at receiver $j$ is denoted by $\textbf{z}_j^\text{II}$, $\textbf{u}_2 = [\textbf{u}_2[1];\cdots;\textbf{u}_2[\tau_1]]  \in \mathbb{C}^{\tau_1M}$, and $\textbf{H}_{2,j}^\text{II} = \text{bd}\{\textbf{H}_{2,j}[\tau_1+1],\cdots,\textbf{H}_{2,j}[2\tau_1]\} \in \mathbb{C}^{\tau_1N \times \tau_1M},j=1,2$.

\underline{\textit{Phase-III}} \textit{(Data Symbol Transmission for Receiver 1 from Two Transmitters)}: This phase spans $\tau_2$ TSs. With the CSI matrices of Phase-I and Phase-II, transmitters 1 and 2 re-construct $\textbf{y}_{1}^\text{I}$ and $\textbf{y}_{1}^\text{II}$, respectively, when the AWGN is ignored. At TS $t=2\tau_1+1\cdots,2\tau_1+\tau_2$, $M$ data symbols (for receiver 1) superposed received AN signals are sent from transmitter $1$,  i.e., $\textbf{x}_1^\text{III}[t] = \textbf{a}_1^{a}[t-2\tau_1] + \phi[t-2\tau_1] \textbf{y}_{1}^\text{I}$. Meanwhile, $M$ data symbols (for receiver 1) superposed received AN signals are sent from transmitter $2$, i.e., $\textbf{x}_2^\text{III}[t] = \textbf{a}_2[t-2\tau_1] + \phi[t-2\tau_1]  \textbf{y}_{1}^\text{II}$. The holistic transmitted signals for Phase-III are written as
\begin{subequations}
\begin{eqnarray}
&&\textbf{x}_1^\text{III} = \textbf{a}_1^{a} + \Phi \textbf{y}_{1}^\text{I}, \\
&& \textbf{x}_2^\text{III} = \textbf{a}_2 + \Phi \textbf{y}_{1}^\text{II}.
\end{eqnarray}
\end{subequations}
The holistic received signals for Phase-III are written as
\begin{eqnarray}
&& \textbf{y}_j^\text{III} = \textbf{H}_{1,j}^\text{III}\textbf{x}_1^\text{III} + \textbf{H}_{2,j}^\text{III}\textbf{x}_2^\text{III} + \textbf{z}_j^\text{III}, \quad j=1,2,
\end{eqnarray}
where the AWGN signal at receiver $j$ is denoted by $\textbf{z}_j^\text{III}$, $\textbf{a}_1^a = [\textbf{a}_1^a[1];\cdots;\textbf{a}_1^a[\tau_2]]  \in \mathbb{C}^{\tau_2M}$, $\textbf{a}_2 = [\textbf{a}_2[1];\cdots;\textbf{a}_2[\tau_2]]   \in \mathbb{C}^{\tau_2M}$, and $\textbf{H}_{i,j}^\text{III} = \text{bd}\{\textbf{H}_{i,j}[2\tau_1+1],\cdots,\textbf{H}_{i,j}[2\tau_1+\tau_2]\}   \in \mathbb{C}^{\tau_2N \times \tau_2M}, i,j=1,2$.

\underline{\textit{Phase-IV}} \textit{(Data Symbol Transmission for Receiver 1 from Transmitter 1)}: This phase spans $\tau_3$ TSs. 
With the CSI matrices of Phase-III, the transmitter 2 re-constructs $\textbf{H}_{2,2}^\text{III}\textbf{x}_2^\text{III}$. At TS $t=2\tau_1+\tau_2+1,\cdots,2\tau_1+\tau_2+\tau_3$, $M$ data symbols (for receiver 1) superposed received AN signals are sent from transmitter $1$, i.e., $\textbf{x}_1^\text{IV}[t] = \textbf{a}_1^{b}[t-2\tau_1-\tau_2] + \omega[t-2\tau_1-\tau_2]\textbf{y}_{1}^\text{I}$. Meanwhile, the transmitter 2 sends $\textbf{x}_2^\text{IV}[t] = \gamma[t-2\tau_1-\tau_2] \textbf{H}_{2,2}^\text{III}\textbf{x}_2^\text{III}$.  The holistic transmitted signals for Phase-IV are written as
 \begin{subequations}
 	\begin{eqnarray}
 		&& \textbf{x}_1^\text{IV} = \textbf{a}_1^{b} + \Omega \textbf{y}_{1}^\text{I},\\
 		&& \textbf{x}_2^\text{IV} = \Gamma \textbf{H}_{2,2}^\text{III}\textbf{x}_2^\text{III},
 	\end{eqnarray}
 \end{subequations}
The holistic received signals for Phase-IV are written as
\begin{eqnarray}
&& \textbf{y}_j^\text{IV} = \textbf{H}_{1,j}^\text{IV}\textbf{x}_1^\text{IV} +
 \textbf{H}_{2,j}^\text{IV}\textbf{x}_2^\text{IV}  + \textbf{z}_j^\text{IV}, \quad j=1,2, 
\end{eqnarray}
where the AWGN signal at receiver $j$ is denoted by $\textbf{z}_j^\text{IV}$, $\textbf{a}_1^b = [\textbf{a}_1^b[1];\cdots;\textbf{a}_1^b[\tau_3]]   \in \mathbb{C}^{\tau_3M}$, and $\textbf{H}_{i,j}^\text{IV}=\text{bd}\{\textbf{H}_{i,j}[2\tau_1+\tau_2+1],\cdots,\textbf{H}_{i,j}[2\tau_1+\tau_2+\tau_3]\}   \in \mathbb{C}^{\tau_3N \times \tau_3M}, i,j=1,2$.

\underline{\textit{Phase-V}}  \textit{(Data Symbol Transmission for Receiver 2 from Two Transmitters)}: This phase is the same as Phase-III, except the role of the transmitters 1 and 2 is swapped. Thus, this phase spans $\tau_2$ TSs as well. With the CSI matrices of Phase-I and Phase-II, transmitters 1 and 2 re-construct $\textbf{y}_{2}^\text{I}$ and $\textbf{y}_{2}^\text{II}$, respectively, when the AWGN is ignored. At TS $t=2\tau_1+\tau_2+\tau_3+1,\cdots,2\tau_1+2\tau_2+\tau_3$, $M$ data symbols (for receiver 2) superposed received AN signals are sent from transmitter $1$, i.e., $\textbf{x}_1^\text{V}[t] = \textbf{b}_1[t-2\tau_1-\tau_2-\tau_3] + \phi[t-2\tau_1-\tau_2-\tau_3] \textbf{y}_{2}^\text{I}$.  Meanwhile, $M$ data symbols (for receiver 2) superposed received AN signals are sent from transmitter $2$, i.e., $\textbf{x}_2^\text{V}[t] = \textbf{b}_2^{a}[t-2\tau_1-\tau_2-\tau_3] + \phi[t-2\tau_1-\tau_2-\tau_3]  \textbf{y}_{2}^\text{II}$. The holistic transmitted signals for Phase-V are written as
\begin{subequations}
\begin{eqnarray}
&&\textbf{x}_1^\text{V} = \textbf{b}_1 + \Phi \textbf{y}_{2}^\text{I}, \\
&& \textbf{x}_2^\text{V} = \textbf{b}_2^{a} + \Phi \textbf{y}_{2}^\text{II},
\end{eqnarray}
\end{subequations}
The holistic received signals for Phase-V are written as
\begin{eqnarray}
&& \textbf{y}_j^\text{V} = \textbf{H}_{1,j}^\text{V}\textbf{x}_1^\text{V} + \textbf{H}_{2,j}^\text{V}\textbf{x}_2^\text{V} + \textbf{z}_j^\text{V}, \quad j=1,2,
\end{eqnarray}
where the AWGN signal at receiver $j$ is denoted by $\textbf{z}_j^\text{V}[t]$, $\textbf{b}_1 = [\textbf{b}_1[1];\cdots;\textbf{b}_1[\tau_2]] \in  \mathbb{C}^{\tau_2M}$, $\textbf{b}_2^a = [\textbf{b}_2^a[1];\cdots;\textbf{b}_2^a[\tau_2]] \in  \mathbb{C}^{\tau_2M}$, and  $\textbf{H}_{i,j}^\text{V}=\text{bd}\{\textbf{H}_{i,j}[2\tau_1+\tau_2+\tau_3+1],\cdots,\textbf{H}_{i,j}[2\tau_1+2\tau_2+\tau_3]\} \in  \mathbb{C}^{\tau_2N \times \tau_2M}, i,j=1,2$.

\underline{\textit{Phase-VI}} \textit{(Data Symbol Transmission for Receiver 2 from Transmitter 2)}: This phase is the same as Phase-IV, except the role of the transmitters 1 and 2 is swapped.  Hence, this phase spans $\tau_3$ TSs as well.  With the CSI matrices of Phase-V,  the transmitter 1 re-constructs $\textbf{H}_{1,1}^\text{V}\textbf{x}_1^\text{V}$. At TS $t=2\tau_1+2\tau_2+\tau_3+1,\cdots,2\tau_1+2\tau_2+2\tau_3$, $M$ data symbols for receiver 2 are sent from transmitter $2$, i.e., $\textbf{x}_1^\text{VI}[t] = \textbf{b}_2^{b}[t- 2\tau_1-2\tau_2-\tau_3]  + \omega[t-2\tau_1-2\tau_2-\tau_3]  \textbf{y}_{2}^\text{II}$. Meanwhile, the transmitter 1 sends $\textbf{x}_2^\text{VI}[t] = \gamma[t-2\tau_1-2\tau_2-\tau_3] \textbf{H}_{1,1}^\text{V}\textbf{x}_1^\text{V}$. The holistic transmitted signals for Phase-VI are written as
\begin{subequations}
	\begin{eqnarray}
		&& \textbf{x}_1^\text{VI} = \Gamma \textbf{H}_{1,1}^\text{V}\textbf{x}_1^\text{V},\\
		&& \textbf{x}_2^\text{VI} =  \textbf{b}_2^{b}  + \Omega  \textbf{y}_{2}^\text{II},
	\end{eqnarray}
\end{subequations}
The holistic received signals for Phase-VI are written as
\begin{eqnarray}
&& \textbf{y}_j^\text{VI} = \textbf{H}_{1,j}^\text{VI}\textbf{x}_1^\text{VI}  +
 \textbf{H}_{2,j}^\text{VI}\textbf{x}_2^\text{VI}+ \textbf{z}_j^\text{VI},\quad j=1,2, 
\end{eqnarray}
where the AWGN signal at receiver $j$ is denoted by $\textbf{z}_j^\text{VI}$, $\textbf{b}_2^b = [\textbf{b}_2^b[1];\cdots;\textbf{b}_2^b[\tau_3]] \in \mathbb{C}^{\tau_3M}$, and $\textbf{H}_{i,j}^\text{VI}=\text{bd}\{\textbf{H}_{i,j}[2\tau_1+2\tau_2+\tau_3+1],\cdots,\textbf{H}_{i,j}[2\tau_1+2\tau_2+2\tau_3]\} \in \mathbb{C}^{\tau_3N \times \tau_3M}, i,j=1,2$.




\underline{\textit{Phase-VII}} \textit{(Interference Recurrence)}: This phase spans $\tau_4$ TSs, which is used to re-transmit the combination of previous interference signals. This re-transmission will not incur new interference, but create useful equations for decoding.  With the CSI matrices of Phase-III to Phase-VI, the transmitter 1 re-constructs $\textbf{H}_{2,2}^\text{IV}\Gamma\textbf{H}_{1,2}^\text{III}\textbf{x}_1^\text{III} -  \textbf{H}_{1,2}^\text{IV}\textbf{x}_1^\text{IV}$, and the transmitter 2 re-constructs $\textbf{H}_{1,1}^\text{VI}\Gamma\textbf{H}_{2,1}^\text{V}\textbf{x}_2^\text{V} -  \textbf{H}_{2,1}^\text{VI}\textbf{x}_2^\text{VI}$. At TS $t=2\tau_1+2\tau_2+2\tau_3+1,\cdots,2\tau_1+2\tau_2+2\tau_3+\tau_4$, the transmitter 1 sends  $\textbf{x}_1^\text{VII}[t] = \theta[t-2\tau_1-2\tau_2-2\tau_3](\textbf{H}_{2,2}^\text{IV}\Gamma
\textbf{H}_{1,2}^\text{III}\textbf{x}_1^\text{III} -  \textbf{H}_{1,2}^\text{IV}\textbf{x}_1^\text{IV})$, and the transmitter 2 sends $\textbf{x}_2^\text{VII}[t] = \theta[t-2\tau_1-2\tau_2-2\tau_3](\textbf{H}_{1,1}^\text{VI}\Gamma\textbf{H}_{2,1}^\text{V}\textbf{x}_2^\text{V} -  \textbf{H}_{2,1}^\text{VI}\textbf{x}_2^\text{VI})$, with $N$ antennas. The holistic transmitted signals for Phase-VII are written as
\begin{subequations}
\begin{eqnarray}
&& \textbf{x}_1^\text{VII} = \Theta  (\textbf{H}_{2,2}^\text{IV}\Gamma\textbf{H}_{1,2}^\text{III}\textbf{x}_1^\text{III} -  \textbf{H}_{1,2}^\text{IV}\textbf{x}_1^\text{IV}),\\
 &&\textbf{x}_2^\text{VII} =  \Theta (\textbf{H}_{1,1}^\text{VI}\Gamma\textbf{H}_{2,1}^\text{V}\textbf{x}_2^\text{V} -  \textbf{H}_{2,1}^\text{VI}\textbf{x}_2^\text{VI}).
\end{eqnarray}
\end{subequations}
The holistic received signals for Phase-VII are written as
\begin{eqnarray}
&& \textbf{y}_j^\text{VII} = \textbf{H}_{1,j}^\text{VII}\textbf{x}_1^\text{VII} + \textbf{H}_{2,j}^\text{VII}\textbf{x}_2^\text{VII} + \textbf{z}_j^\text{VII}, \quad j=1,2,
\end{eqnarray}
where the AWGN signal at receiver $j$ is denoted by $\textbf{z}_j^\text{VII}$, and  $\textbf{H}_{i,j}^\text{VII}=\text{bd}\{\textbf{H}_{i,j}[2\tau_1+2\tau_2+2\tau_3+1],\cdots,\textbf{H}_{i,j}[2\tau_1+2\tau_2+2\tau_3+\tau_4]\} \in \mathbb{C}^{\tau_4N \times \tau_4N}, i,j=1,2$.

\begin{figure*} 
	\begin{eqnarray} \label{H1}
		&&  \begin{bmatrix}
			\textbf{y}_1^\text{III}    \\
			\textbf{y}_1^\text{IV}    \\
			\textbf{y}_1^\text{VII} - \textbf{H}_{2,1}^\text{VII}\Theta(\textbf{H}_{1,1}^\text{VI}\Gamma\textbf{y}_1^\text{V} - \textbf{y}_1^\text{VI}) 
		\end{bmatrix}   = 
		\underbrace{\begin{bmatrix}
				\textbf{H}_{1,1}^\text{III} & \textbf{0} & 	\textbf{H}_{2,1}^\text{III} \\
				\textbf{0} & 	\textbf{H}_{1,1}^\text{IV} & \textbf{H}_{2,1}^\text{IV} \Gamma \textbf{H}_{2,2}^\text{III}\\
				\textbf{H}_{1,1}^\text{VII} \Theta \textbf{H}_{2,2}^\text{IV} \Gamma \textbf{H}_{1,2}^\text{III} & -\textbf{H}_{1,1}^\text{VII}\Theta\textbf{H}_{1,2}^\text{IV} & \textbf{0}
		\end{bmatrix}}_{\textbf{H}_1} 	\begin{bmatrix}
			\textbf{a}_1^a \\
			\textbf{a}_1^b \\
			\textbf{a}_2
		\end{bmatrix} \nonumber \\
		&& + \begin{bmatrix}
			\textbf{H}_{1,1}^\text{III}\Phi  & \textbf{H}_{2,1}^\text{III}\Phi \\
			\textbf{H}_{1,1}^\text{IV} \Omega & \textbf{H}_{2,1}^\text{IV} \Gamma \textbf{H}_{2,2}^\text{III}\Phi  \\
			\textbf{H}_{1,1}^\text{VII}\Theta(\textbf{H}_{2,2}^\text{IV}\Gamma\textbf{H}_{1,2}^\text{III}\Phi -  \textbf{H}_{1,2}^\text{IV}\Omega) & \textbf{0}
		\end{bmatrix}\begin{bmatrix}
			\textbf{y}_1^\text{I} \\
			\textbf{y}_1^\text{II}
		\end{bmatrix} + \underline{\textbf{z}}_1.
	\end{eqnarray}
	\hrule
		\begin{eqnarray}
		&& I(\textbf{b}_2^a,\textbf{b}_2^b,\textbf{b}_1;\textbf{y}_1|\textbf{a}_1^a,\textbf{a}_1^b,\textbf{a}_2) \overset{(a)}{\le}  I(\textbf{H}^{\text{I}}_{1,1} \textbf{u}_1, \textbf{H}^{\text{II}}_{2,1} \textbf{u}_2, \textbf{H}_{1,1}^\text{V}(\textbf{b}_1 +  \Phi \textbf{H}_{1,2}^\text{I}\textbf{u}_1) + \textbf{H}_{2,1}^\text{V} (\textbf{b}_2^{a}+ \Phi  \textbf{H}_{2,2}^\text{II}\textbf{u}_2), \nonumber \\
		&&   \textbf{H}_{1,1}^\text{VI}\Gamma\textbf{H}_{1,1}^\text{V}(\textbf{b}_1 +   \Phi \textbf{H}_{1,2}^\text{I}\textbf{u}_1) +
		\textbf{H}_{2,1}^\text{VI} (\textbf{b}_2^{b} + \Omega  \textbf{H}_{2,2}^\text{II}\textbf{u}_2);\textbf{y}_1|\textbf{a}_1^a,\textbf{a}_1^b,\textbf{a}_2)  - I(\textbf{u};\textbf{y}_1|  \textbf{b}_2^a,\textbf{b}_2^b,\textbf{b}_1,\textbf{a}_1^a,\textbf{a}_1^b,\textbf{a}_2) \nonumber \\
		&& \underset{\text{SNR} \rightarrow {\cal{1}}}{\overset{(b)}{=}} \text{rank}  \left\{
		\underbrace{\begin{bmatrix}
				\textbf{I}_{N\tau_1} &  \textbf{0} & \textbf{0} & \textbf{0}\\
				\textbf{0} & 	\textbf{I}_{N\tau_1} & \textbf{0} & \textbf{0} \\
				\textbf{H}_{1,1}^\text{III}\Phi & \textbf{H}_{2,1}^\text{III}\Phi & \textbf{0} & \textbf{0}\\
				\textbf{H}_{1,1}^\text{IV}\Omega & \textbf{H}_{2,1}^\text{IV}\Gamma\textbf{H}_{2,2}^\text{III}\Phi & \textbf{0} & \textbf{0}\\
				\textbf{0} &  \textbf{0}  & \textbf{I}_{N\tau_2} & \textbf{0} \\
				\textbf{0} & \textbf{0} &  \textbf{0} & \textbf{I}_{N\tau_3} \\
				\textbf{H}_{1,1}^\text{VII}\Theta(\textbf{H}_{2,2}^\text{IV}\Gamma\textbf{H}_{1,2}^\text{III}\Phi - \textbf{H}_{1,2}^\text{IV}\Omega) & \textbf{0} & \textbf{H}_{2,1}^\text{VII} \Theta \textbf{H}_{1,1}^\text{VI} \Gamma &  -\textbf{H}_{2,1}^\text{VII} \Theta
		\end{bmatrix}}_{\textbf{A}}\right\} \log \text{SNR} \nonumber \\
		&&  - \text{rank} \left\{\underbrace{\begin{bmatrix}
				\textbf{H}_{1,1}^\text{I} &  \textbf{0} \\
				\textbf{0} & 		\textbf{H}_{2,1}^\text{II}   \\
				\textbf{H}_{1,1}^\text{III}\Phi\textbf{H}_{1,1}^\text{I} &  \textbf{H}_{2,1}^\text{III}\Phi\textbf{H}_{2,1}^\text{II}  \\
				\textbf{H}_{1,1}^\text{IV}\Omega\textbf{H}_{1,1}^\text{I} & \textbf{H}_{2,1}^\text{IV}\Gamma\textbf{H}_{2,2}^\text{III}\Phi\textbf{H}_{2,1}^\text{II}   \\
				\textbf{H}_{1,1}^\text{V}\Phi\textbf{H}_{1,2}^\text{I} & \textbf{H}_{2,1}^\text{V}\Phi\textbf{H}_{2,2}^\text{II}  \\
				\textbf{H}_{1,1}^\text{VI}\Gamma\textbf{H}_{1,1}^\text{V}\Phi\textbf{H}_{1,2}^\text{I} & 	\textbf{H}_{2,1}^\text{VI}\Omega\textbf{H}_{2,2}^\text{II}   \\
				\textbf{H}_{1,1}^\text{VII}\Theta(\textbf{H}_{2,2}^\text{IV}\Gamma\textbf{H}_{1,2}^\text{III}\Phi-\textbf{H}_{1,2}^\text{IV}\Omega)\textbf{H}_{1,1}^\text{I} & 	\textbf{H}_{2,1}^\text{VII}\Theta(\textbf{H}_{1,1}^\text{VI}\Gamma\textbf{H}_{2,1}^\text{V}\Phi-\textbf{H}_{2,1}^\text{VI}\Omega)\textbf{H}_{2,2}^\text{II} 
		\end{bmatrix}}_{\textbf{B}}\right\} \log \text{SNR} \nonumber \\
		&& \overset{(c)}{=} N(2\tau_1 + \tau_2 + \tau_3)\log \text{SNR} - \min\{N(2\tau_1 + \min\{\tau_1,\tau_2\} + \min\{\tau_1,\tau_3\}),2M\tau_1\} \log \text{SNR}.  \label{Q1}
	\end{eqnarray} 
	\hrule
\end{figure*}

For decoding, due to the symmetry, we only need to perform analysis at one receiver. The final decoding equation at receiver 1 is given  in \eqref{H1}, where the AWGN signal is denoted by $\underline{\textbf{z}}_1$. The decoding of data symbols  is only related to $\textbf{H}_1$, since the impact of $\textbf{y}_1^\text{I}$ and $\textbf{y}_1^\text{II}$ can be removed. The rank of $\textbf{H}_1$ in \eqref{H1} is $\min\{N(\tau_2+\tau_3+ \min\{\tau_3,\tau_4\}), M(2\tau_2+\tau_3)\}$, whose reason is given in Appendix A.  Since the impact of $\textbf{y}_1^\text{I}$ and $\textbf{y}_1^\text{II}$ is removed for decoding, the optimal $\tau_2^*,\tau_3^*,\tau_4^*$ can be found in \cite{64}, which is given by 
$(\tau_2^*,\tau_3^*,\tau_4^*) = (2N-M,2M-N,2M-N)$. It can be verified that the rank of $\textbf{H}_1$ is equal to the number of data symbols for receiver 1, i.e., $\min\{N(\tau_2^*+\tau_3^*+ \min\{\tau_3^*,\tau_4^*\}), M(2\tau_2^*+\tau_3^*)\} = M(2\tau_2^*+\tau_3^*)$.

For security, due to the symmetry, we only need to perform analysis at one receiver. Given the notations $\textbf{y}_1 = [\textbf{y}_1^\text{I};\cdots;\textbf{y}_1^\text{VII}]$ and $\textbf{u} = [\textbf{u}_1;\textbf{u}_2]$, the  information leakage $I(\textbf{b}_2^a,\textbf{b}_2^b,\textbf{b}_1;\textbf{y}_1|\textbf{a}_1^a,\textbf{a}_1^b,\textbf{a}_2)$ is calculated in \eqref{Q1}, where the reason of each step is given as follows:
\begin{enumerate}[(a)]
	\item $I(\textbf{b}_2^a,\textbf{b}_2^b,\textbf{b}_1,\textbf{u};\textbf{y}_1|\textbf{a}_1^a,\textbf{a}_1^b,\textbf{a}_2)$ =  $I(\textbf{b}_2^a,\textbf{b}_2^b,\textbf{b}_1;\textbf{y}_1|\textbf{a}_1^a,\textbf{a}_1^b,\textbf{a}_2)$ + $I(\textbf{u};\textbf{y}_1|  \textbf{b}_2^a,\textbf{b}_2^b,\textbf{b}_1,\textbf{a}_1^a,\textbf{a}_1^b,\textbf{a}_2)$, and applying the data processing inequality for the Markov chain $(\textbf{b}_2^{a},  \textbf{b}_2^{b}, \textbf{b}_1,\textbf{u}) \rightarrow (\textbf{H}^{\text{I}}_{1,1} \textbf{u}_1, \textbf{H}^{\text{II}}_{2,1} \textbf{u}_2, \textbf{H}_{1,1}^\text{V}(\textbf{b}_1 +  \Phi \textbf{H}_{1,2}^\text{I}\textbf{u}_1) + \textbf{H}_{2,1}^\text{V} (\textbf{b}_2^{a}+ \Phi  \textbf{H}_{2,2}^\text{II}\textbf{u}_2),\textbf{H}_{1,1}^\text{VI}\Gamma\textbf{H}_{1,1}^\text{V}(\textbf{b}_1 +   \Phi \textbf{H}_{1,2}^\text{I}\textbf{u}_1) +
	\textbf{H}_{2,1}^\text{VI} (\textbf{b}_2^{b} + \Omega  \textbf{H}_{2,2}^\text{II}\textbf{u}_2)) \rightarrow \textbf{y}_1$.
	\item When input is circularly symmetric complex Gaussian,
	according to \cite{100}, rewriting into $\log \text{det}(I + \text{SNR}\textbf{AA}^H) -	\log \text{det}(\textbf{I} + \text{SNR}\textbf{BB}^H)$, and using Lemma 2 in \cite{31}.
	\item It can be verified by Gaussian elimination that the rank of matrix \textbf{A} is $N(2\tau_1 + \tau_2 + \tau_3)$. The rank of matrix
	\textbf{B} is $\min\{N(2\tau_1 + \min\{\tau_1,\tau_2\} + \min\{\tau_1,\tau_3\}),2M\tau_1\}$, whose reason is given in Appendix B.
\end{enumerate}
Therefore, to ensure $I(\textbf{b}_2^a,\textbf{b}_2^b,\textbf{b}_1;\textbf{y}_1|\textbf{a}_1^a,\textbf{a}_1^b,\textbf{a}_2) = o(\log \text{SNR})$, according to \eqref{Q1}, $\tau_1$ should  follow that
\begin{subequations}
\begin{eqnarray} 
&& \tau_2 \le \tau_1, \label{1S} \\
&& \tau_3 \le \tau_1, \label{2S} \\
&& N(2\tau_1 + \tau_2 + \tau_3) \le 2M\tau_1, \label{3S}
\end{eqnarray}
\end{subequations}
Then, substituting the $(\tau_2^*,\tau_3^*,\tau_4^*) = (2N-M,2M-N,2M-N)$ into \eqref{1S}-\eqref{3S} and simplifying the expression, we have
	\begin{eqnarray}
	\max\left \{\frac{N(M+N)}{2(M-N)},2M-N \right\} \le \tau_1. 
	\end{eqnarray}
To maximize the sum-SDoF lower bound achieved by our scheme, $\tau_1$ should be as small as possible. This is because, Phase-I and Phase-II do not contain any fresh data symbols. Consequently, the optimal $\tau_1^*$ is given by
\begin{equation} \label{O2}
\tau_1^* = \begin{cases}
\frac{N(M+N)}{2(M-N)}, & N < M \le \frac{7+\sqrt{33}}{8}N, \\
2M-N, & \frac{7+\sqrt{33}}{8}N < M \le 2N.
\end{cases}
\end{equation}
Our scheme has delivered $2M(2\tau_2+\tau_3)$ data symbols over $2\tau_1+2\tau_2+2\tau_3+\tau_4$ TSs. With the above $(\tau_1^*,\tau_2^*,\tau_3^*,\tau_4^*)$, the sum-SDoF lower bound in \eqref{LB} for $N< M \le 2N$ is achieved.

\subsection{$2N < M$: Adopt the Transmission Scheme in \cite{35}}

 Intuitively, since the number of useful equations at the two receivers is at most $2N$ per TS, the data symbols cannot be decoded by interference recurrence if we send more than $2N$ data symbols per TS. This motivates us to send $2N$ data symbols from one transmitter for one receiver at one TS, as the scheme in \cite{35} does, where the lower bound is $4N/5$.

\section{Conclusions}

We have obtained a sum-SDoF lower bound of the MIMO XCCM with delayed CSIT by proposing a transmission scheme. This transmission scheme can be deemed as a generalized version of the scheme in \cite{64} for symmetric antenna configurations. We have derived the optimal phase duration for AN transmission based on security analysis. In the future,  the research can be devoted to: 1) Finding a
linear sum-SDoF upper bound; 2) Extending the proposed scheme to the one for arbitrary antenna configurations with the absence of symmetry.

\section*{Appendix}

\subsection{Rank Analysis for Matrix $\textbf{H}_1$}
 
 The rank of  $\textbf{H}_1$ is equal to the sum of the rank of
 \begin{equation}
\textbf{L} = \begin{bmatrix}
	\textbf{H}_{1,1}^\text{III} & \textbf{0} & 	\textbf{H}_{2,1}^\text{III} \\
	\textbf{0} & 	\textbf{H}_{1,1}^\text{IV} & \textbf{H}_{2,1}^\text{IV} \Gamma \textbf{H}_{2,2}^\text{III}
\end{bmatrix},\nonumber
 \end{equation}
 and the rank of
  \begin{equation}
 \textbf{U} =  \begin{bmatrix}\textbf{H}_{1,1}^\text{VII} \Theta \textbf{H}_{2,2}^\text{IV} \Gamma \textbf{H}_{1,2}^\text{III}& -\textbf{H}_{1,1}^\text{VII}\Theta\textbf{H}_{1,2}^\text{IV} & \textbf{0}\end{bmatrix}. \nonumber
 \end{equation}

Due to the linear independence, the rank of $\textbf{L}$ is equal to the sum of the rank of sub-matrix $[\textbf{H}_{1,1}^\text{III}, \textbf{0}, 	\textbf{H}_{2,1}^\text{III} ]$ and the rank of sub-matrix $[\textbf{0}, \textbf{H}_{1,1}^\text{IV}, \textbf{H}_{2,1}^\text{IV} \Gamma \textbf{H}_{2,2}^\text{III}]$. The rank of sub-matrix $[	\textbf{H}_{1,1}^\text{III}, \textbf{0}, 	\textbf{H}_{2,1}^\text{III} ]$ is $N\tau_2$. 
On the other hand, when $N < M$, the rank of $\textbf{H}_{2,1}^\text{IV} \Gamma \textbf{H}_{2,2}^\text{III}$ is $N\min\{ \tau_2,\tau_3\}$. Thus, the rank of 
$[\textbf{H}_{1,1}^\text{IV}, \textbf{H}_{2,1}^\text{IV} \Gamma \textbf{H}_{2,2}^\text{III}]$ is $N\tau_3$. Consequently, the rank of
$\textbf{L}$ is $N(\tau_2+\tau_3)$. The rank of $\textbf{U}$ is determined by the sub-matrix $[\textbf{H}_{1,1}^\text{VII} \Theta \textbf{H}_{2,2}^\text{IV} \Gamma \textbf{H}_{1,2}^\text{III},-\textbf{H}_{1,1}^\text{VII}\Theta\textbf{H}_{1,2}^\text{IV}]$, which can be decomposed into the multiplication of $\textbf{H}_{1,1}^\text{VII} \Theta$ and $[ \textbf{H}_{2,2}^\text{IV} \Gamma \textbf{H}_{1,2}^\text{III},-\textbf{H}_{1,2}^\text{IV}]$. When $N < M$, the rank of $\textbf{H}_{1,1}^\text{VII} \Theta$ is $N\min\{\tau_4,\tau_3\}$ and the rank of $[ \textbf{H}_{2,2}^\text{IV} \Gamma \textbf{H}_{1,2}^\text{III},-\textbf{H}_{1,2}^\text{IV}]$ is $N\tau_3$. Since the rank of multiplication of two matrices is determined by the minimal rank of one of them, thus the rank of $\textbf{U}$ is $N\min\{ \tau_3,\tau_4\}$. Therefore,  we conclude that the rank of $\textbf{H}_1$ is  $N(\tau_2+\tau_3 + \min\{\tau_3,\tau_4\})$.

\subsection{Rank Analysis for Matrix $\textbf{B}$}

For matrix \textbf{B}, the blocks $\textbf{H}_{1,1}^\text{III}\Phi\textbf{H}_{1,1}^\text{I} $, $\textbf{H}_{1,1}^\text{IV}\Omega\textbf{H}_{1,1}^\text{I}$, and $	\textbf{H}_{1,1}^\text{VII}\Theta(\textbf{H}_{2,2}^\text{IV}\Gamma\textbf{H}_{1,2}^\text{III}\Phi-\textbf{H}_{1,2}^\text{IV}\Omega)\textbf{H}_{1,1}^\text{I}$ are generated from $\textbf{H}_{1,1}^\text{I}$, the blocks $\textbf{H}_{2,1}^\text{III}\Phi\textbf{H}_{2,1}^\text{II}$ and $\textbf{H}_{2,1}^\text{IV}\Gamma\textbf{H}_{2,2}^\text{III}\Phi\textbf{H}_{2,1}^\text{II}$ are generated from $\textbf{H}_{2,1}^\text{II}$, the  block $\textbf{H}_{2,1}^\text{VII}\Theta(\textbf{H}_{1,1}^\text{VI}\Gamma\textbf{H}_{2,1}^\text{V}\Phi-\textbf{H}_{2,1}^\text{VI}\Omega)\textbf{H}_{2,2}^\text{II}$ is generated from $\textbf{H}_{2,1}^\text{V}\Phi\textbf{H}_{2,2}^\text{II}$ and $\textbf{H}_{2,1}^\text{VI}\Omega\textbf{H}_{2,2}^\text{II}$. Therefore, the rank of $\textbf{B}$ is equivalent to the rank of the following matrix:
 \begin{equation}
\begin{bmatrix}
\textbf{H}_{1,1}^\text{I} &  \textbf{0} \\
\textbf{0} & 		\textbf{H}_{2,1}^\text{II}   \\
 
\textbf{H}_{1,1}^\text{V}\Phi\textbf{H}_{1,2}^\text{I} & \textbf{H}_{2,1}^\text{V}\Phi\textbf{H}_{2,2}^\text{II}  \\
\textbf{0} & 	\textbf{H}_{2,1}^\text{VI}\Omega\textbf{H}_{2,2}^\text{II}   
\end{bmatrix}.  \nonumber 
\end{equation}
The rank of the above matrix is $\min\{N(2\tau_1 + \min\{\tau_1,\tau_2\} + \min\{\tau_1,\tau_3\}),2M\tau_1\}$, since the ranks of  $\textbf{H}_{1,1}^\text{V}\Phi\textbf{H}_{1,2}^\text{I}$ and $\textbf{H}_{2,1}^\text{V}\Phi\textbf{H}_{2,2}^\text{II}$ are $N\min\{\tau_1,\tau_2\}$, the rank of $\textbf{H}_{2,1}^\text{VI}\Omega\textbf{H}_{2,2}^\text{II}$ is $N\min\{\tau_1,\tau_3\}$, the ranks of $\textbf{H}_{1,1}^\text{I}$ and $\textbf{H}_{2,1}^\text{II}$ are $N\tau_1$. As a result, we conclude that the rank of $\textbf{B}$ is $\min\{N(2\tau_1 + \min\{\tau_1,\tau_2\} + \min\{\tau_1,\tau_3\}),2M\tau_1\}$.

\bibliographystyle{IEEEtran}
\bibliography{XC}
 
\end{document}